\documentclass{actapoly}

\usepackage{graphics,graphicx}
\usepackage{arydshln} 
\usepackage{xcolor}
\usepackage{wasysym}
\usepackage{natbib}
\usepackage[utf8]{inputenc}
\usepackage[T1]{fontenc}
\usepackage{lmodern}

\newcommand{\nh}{\ensuremath{N_\text{H}}}

\begin{document}

\title{X-ray Variability Study of Polar Scattered Seyfert\,1 Galaxies}

\institution{remeis}{Dr. Karl Remeis-Observatory, Universit\"at Erlangen-N\"urnberg \& ECAP, Sternwartstrasse 7, 96049 Bamberg, Germany}
\institution{kadler}{Lehrstuhl für Astronomie, Universit\"at W\"urzburg, Campus Hubland Nord, Emil-Fischer-Straße 31, 97074 W\"urzburg, Germany}
\institution{esac}{European Space Astronomy Centre of ESA, PO Box 78, Villanueva de la Ca\~nada, 28691 Madrid, Spain}
\institution{giovanni}{Centro de Astrobiolog\'ia (CSIC-INTA), Dep. de Astrof\'isica; ESAC, PO Box 78, E-28691 Villanueva de la Ca\~nada, Madrid, Spain}

\correspondingauthor[T. Beuchert]{Tobias Beuchert}{remeis,kadler}{tobias.beuchert@sternwarte.uni-erlangen.de}
\author[J.~Wilms]{J\"orn Wilms}{remeis}
\author[M.~Kadler]{Matthias Kadler}{remeis,kadler}
\author[A.~L.~Longinotti]{Anna Lia Longinotti}{esac}
\author[M.~Guainazzi]{Matteo Guainazzi}{esac}
\author[G.~Miniutti]{Giovanni Miniutti}{giovanni}
\author[I.~de la Calle]{Ignacio de la Calle}{esac}

\begin{abstract}
  We study 12 Seyfert\,1 galaxies with a high level of optical
  polarization. Optical light emerging from the innermost regions is
  predominantly scattered in a polar region above the central engine
  directly in our line of sight. These sources show characteristics of
  Seyfert\,2 galaxies such as, e.g., polarized broad lines. The
  polarization signatures suggest a viewing angle of 45\,$^\circ$
  classifying them as intermediate Seyfert\,1/2 types. The unified
  model predicts this line of sight to pass through the outer layer of
  the torus resulting in significant soft X-ray variability due to a
  strongly varying column density. The aim is to find evidence for
  this geometrical assumption in the spectral variability of all
  available historical observations of these sources by
  \textit{XMM-Newton} and \textit{Swift}.
\end{abstract}


\maketitle

\section{Introduction}
\begin{figure}
  \centering
  \includegraphics[width=0.9\columnwidth]{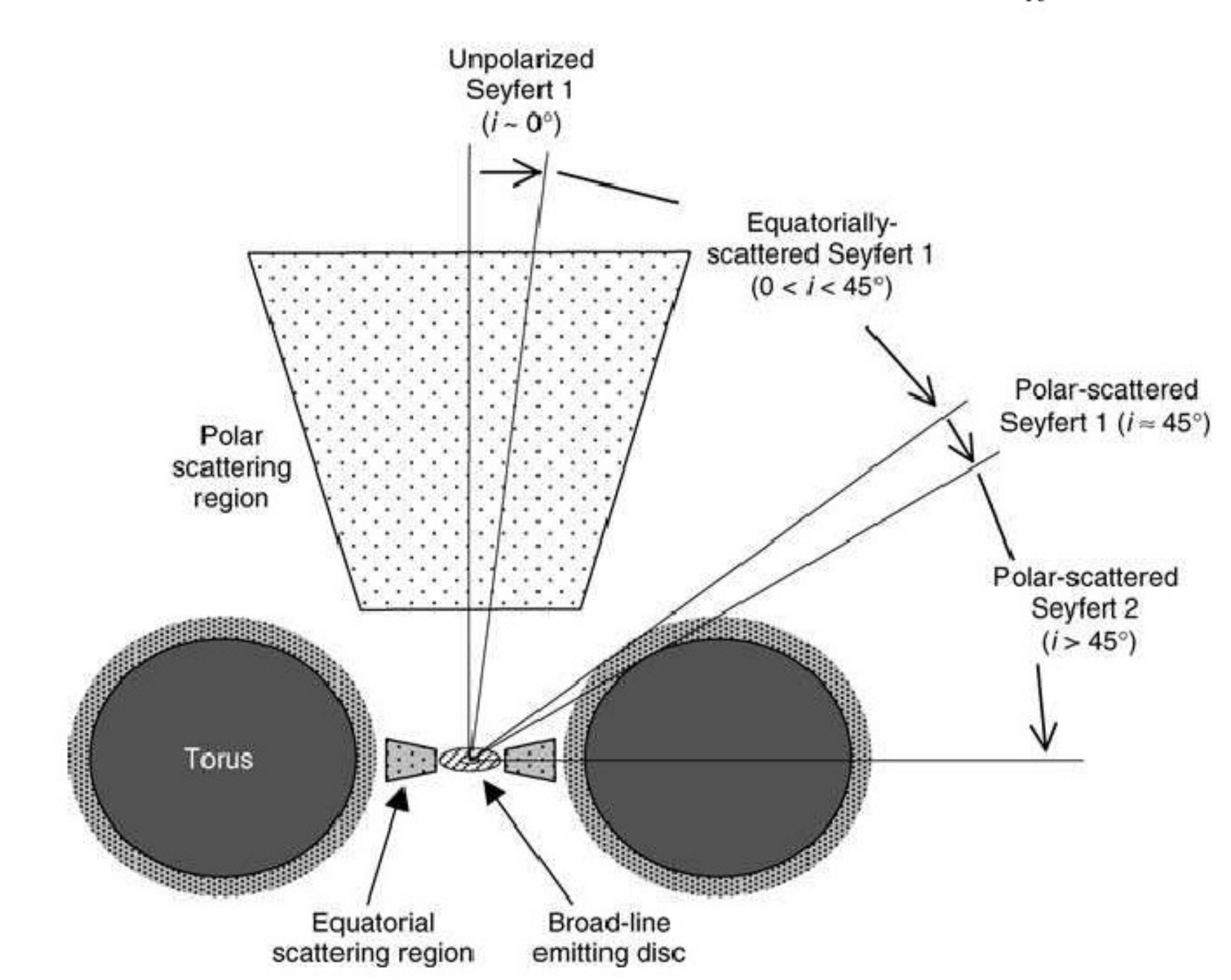}
  \caption{Polar- and equatorial scattering regions according to the
    unified scheme of AGN. Adopted from \cite{2004MNRAS.350..140S}.}
  \label{fig:unifiedmodel}
\end{figure}

According to the unified model of Active Galactic Nuclei/AGN
\cite{Antonucci2012} Seyfert\,1 and Seyfert\,2 galaxies are the same
type of galaxies, but seen under different inclination angles
(Fig.~\ref{fig:unifiedmodel}). At inclinations $\lesssim 45^\circ$,
Seyfert\,1 galaxies are typically either optically unpolarized or
polarized due to predominantly equatorial scattering. Seyfert\,2
galaxies have an inclination of $\gtrsim45^\circ$ and show mainly
optical polarization features due to polar-scattering, as the line of
sight of Seyfert\,2 galaxies passes through the optically thick torus.
This picture usually works well, however, \cite{2004MNRAS.350..140S}
identify 12~Seyfert\,1 galaxies that exhibit optical polarization
similar to Seyfert\,2 galaxies. They conclude that these
polar-scattered Seyfert\,1 galaxies are seen under an inclination of
$i \sim 45^\circ$ and thus represent the transition between unobscured
Seyfert\,1 and obscured Seyfert\,2 galaxies
(Fig.~\ref{fig:unifiedmodel}). The line of sight towards these
galaxies therefore passes through the outer layers of the torus, where
significant absorption is still expected and suppresses polarized
light from the equatorial scattering region but not the
polar-scattered light. We assume these outer layers to be a
non-homogeneous gas and dust medium which might be stripped off by
nuclear radiation resulting in a highly variable column density
towards the observer. X-ray observations of these polar-scattered
Seyfert\,1 galaxies should therefore exhibit a strongly variable \nh.\\
We compare the X-ray properties of a sample of 12 polar-scattered
Seyfert\,1 galaxies and 11 equatorial-scattered Seyfert\,1 galaxies
(see Table~\ref{tab:sourcelist}) for which no absorption variability
is expected according to the line-of-sight. As of now only the first
set of polar-scattered Seyfert\,1 galaxies has been analyzed. In
Table~\ref{tab:sourcelist} the dashed line separates sources with
sufficient signal-to-noise spectra to constrain \nh values (top) from
sources where the spectra do not allow to constrain the column density
(bottom).
\begin{table}
\caption{The sample from Smith et al.: {\it top half:} polar-scattered
  Sy~I galaxies, {\it bottom half:} equatorial scattered Sy~I galaxies
  -- the control sample. The numbers denote the amount of pointings
  with the corresponding satellite.}\label{tab:sourcelist} 

\vspace{0.5cm}

\centering
 \begin{tabular}{lll}
\toprule
 Source Name & {\it Swift} & {\it XMM-Newton} \\
 & XRT & pn \\
\midrule
NGC 3227     & 7 & 2 \\
NGC 4593     & 5 & 2 \\
Mrk 704      & 5 & 2  \\
Fairall 51   & 2 & 2  \\
ESO 323-G077 & 3 & 1  \\
Mrk 1218     & 7 & 2  \\
UGC 7064     & 2 & 1  \\
Mrk 766      & 33&15  \\
\hdashline[1pt/2pt]
Mrk 321      & 1 & 1  \\
Mrk 376      & 2 & \\
IRAS 15091$-$2107 & & 1 \\
Mrk 1239 & & 1  \\
\midrule
Akn 120      & 3 & 1  \\
1Zwl         & 7 & 2   \\
KUV 18217+6419 & 1 & 22  \\
Mrk 006      & 4 & 4  \\
Mrk 304      &   & 2  \\
Mrk 509      & 23 & 17  \\
Mrk 841      & 4 & 5  \\
Mrk 876      & 16 & 2  \\
Mrk 985      & &  \\
NGC 3783     & 6 & 4  \\
NGC 4151     & 4 & 17  \\
\bottomrule
\end{tabular}
\end{table}
When studying the properties of absorption we have to take into
account source intrinsic absorption both of neutral and ionized
matter. The neutral, cold absorption can be due to several spatially
distinct regions of an AGN. Theory
\cite{2001ApJ...561..684K,2008MNRAS.384L..24C} predicts a relatively
cold, dense phase to exists in equilibrium with the partially ionized
outflowing gas forming the broad line region (BLR). These embedded BLR
clouds are commonly believed to be gravitationally bound to the center
of mass. They are good candidates to cause occultation events and an
attenuation of the soft X-ray spectrum. The other possibility farther
out are variable structures at the outer torus layer being possibly
evaporated due to the in-falling radiation. Whereas cold absorbing gas
and dust is assumed to be confined to the obscuring torus, warm,
partially ionized gas -- also contributing to the attenuation of the
soft X-rays as so called warm absorber \cite{1995MNRAS.273.1167R} --
can be most likely combined with extended gas closer to the central
engine than the inner torus boundary flowing outward
\cite{Costantini2010} and ideally fully covering the line-of-sight
\cite{2005A&A...431..111B}.

\section{Methods}
\begin{figure}
  \centering
  \includegraphics[width=0.9\columnwidth]{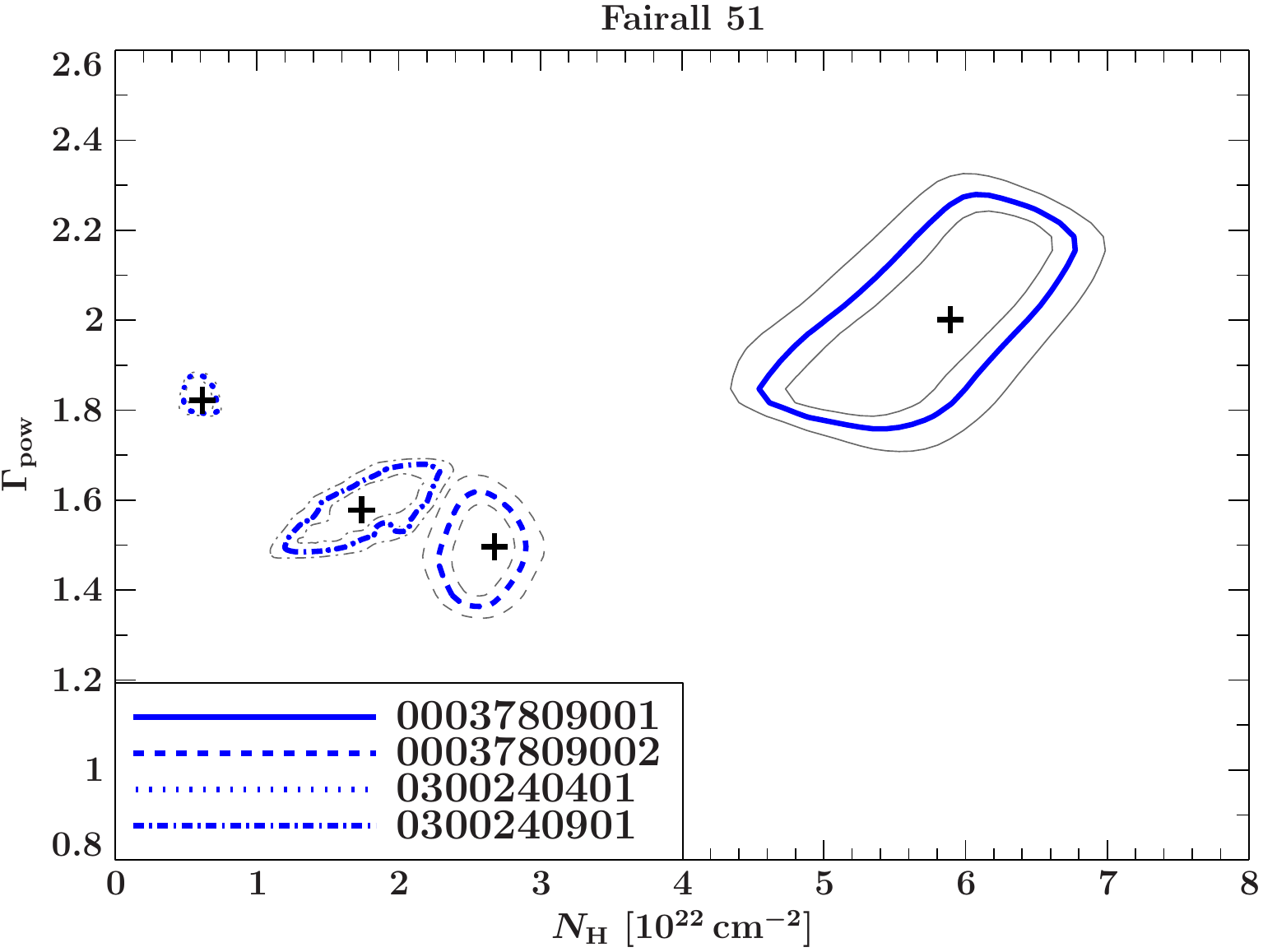}\\
  \includegraphics[width=0.9\columnwidth]{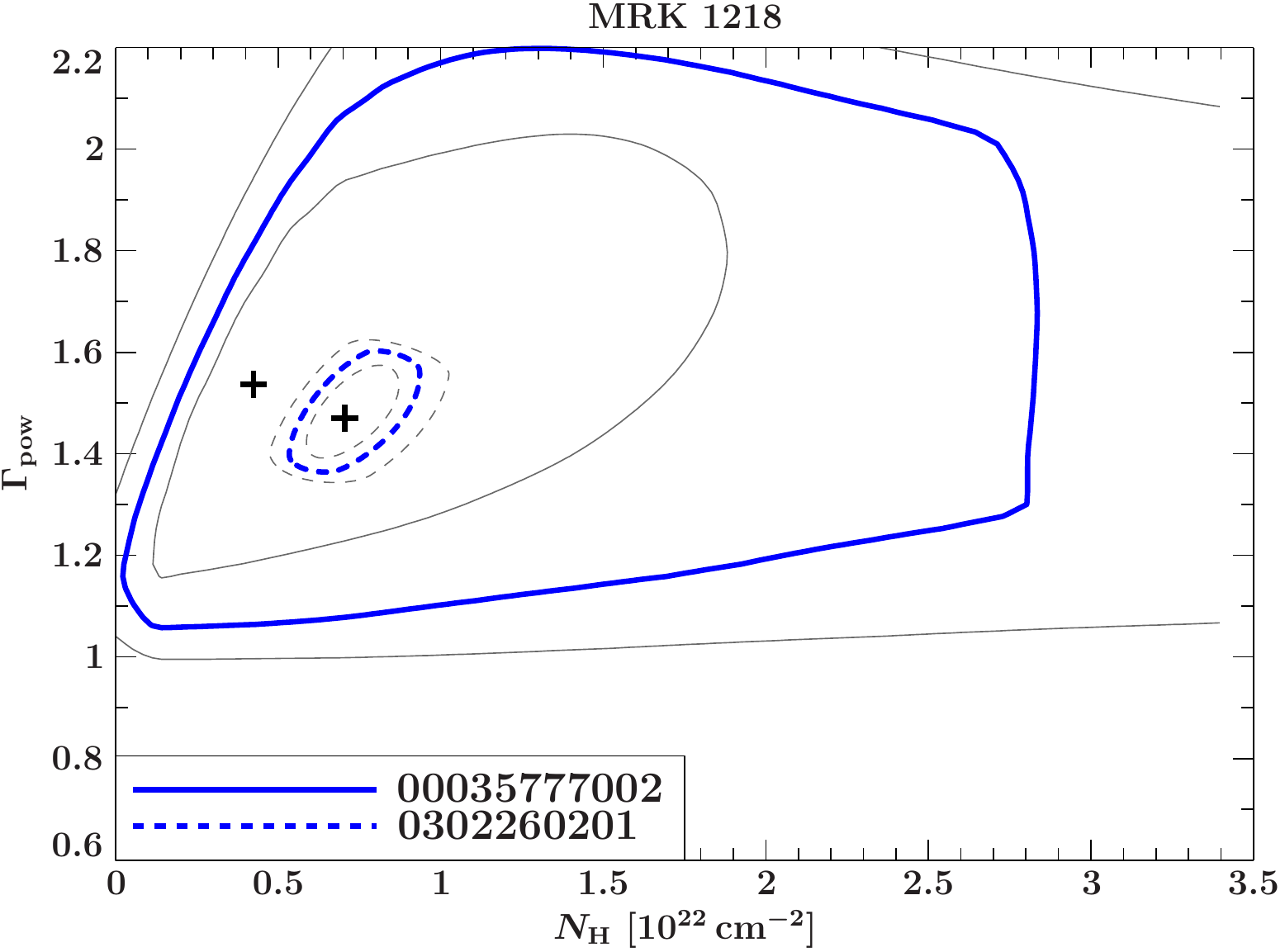}\\
  \caption{Combined contours of all observations of Fairall~51 (top)
    and Mrk~1218 (bottom). The numbers within the plot panels are
    appropriate observation ids.} 
  \label{fig:combcont}
\end{figure}

Here we limit ourselves to the first sample of polar scattered Seyfert
galaxies. The search for variability in these sources requires
consistent model fitting in order to ensure that the detected
variability of absorption indeed originates in the proposed region. An
important task is to disentangle consistently the contributions of
warm- and cold absorption via spectral model fitting in order to
locate the absorber. Distinguishing two absorption components, neutral
and partially ionized, is a challenging task when dealing with low
signal-to-noise spectra, such as {\it Swift} data. {\it XMM-Newton}
observations, however, can help to constrain properties of one or more
warm absorber phases. The picture is getting even more complicated
when considering the warm absorber not to be a homogeneous gas but
rather dynamic and indeed variable both in covering fraction and
column density as shown for Mrk~704 by \cite{2011A&A...533A...1M}. As
a consequence we can not easily draw a conclusion on warm absorber
parameters from one observation to another. As the partially ionized
phase also can attenuate the soft continuum to some extent, we have to
assume that both, ionized and neutral phases do contribute to the
overall continuum absorption within the suggested line-of-sight at
$\sim 45\,^{\circ}$ inclination.\\
Within this work, however, we are explicitly interested in the
variability of neutral absorption. At least for now we do not consider
further the warm absorber contributions unless explicitly necessary.
When searching for variability a consistency check of all modeled
observations is necessary. This is done by examining correlations of
column density related parameters of each fit and observation and
calculating appropriate confidence levels.   
In order to find appropriate model fits, a bottom-up procedure is
chosen starting with a power-law photon continuum,
\begin{equation}
N_\mathrm{ph, full}(E)=A_\mathrm{full}e^{-\sigma{\nh}_\mathrm{int}} (E^{-\Gamma}+\mathcal{G}_{\tt Fe})e^{-\sigma{\nh}_{\mathrm{Gal}}}
\end{equation}
which is fully covered by Galactic and source-intrinsic neutral
matter. We further test a model where the continuum source is
partially covered by the source intrinsic absorber,
\begin{align}
N_\mathrm{ph,partial}(E)=&A_\mathrm{partial} \left[(1-c)\,c\,e^{-\sigma{\nh}_\mathrm{int}}\right]
\nonumber\\
&(E^{-\Gamma}+\mathcal{G}_{\tt Fe})e^{-\sigma{\nh}_\mathrm{gal}}
\end{align}
and finally a partially covered source with an ionized medium in front
of it (a ``warm absorber''),
\begin{align}
N_\mathrm{ph,WA}(E)=&A_\mathrm{WA}\mathrm{WA(1)}\,\mathrm{WA(2)} 
\nonumber\\
&e^{-\sigma{\nh}_\mathrm{int}}
(E^{-\Gamma}+\mathcal{G}_{\tt Fe})e^{-\sigma{\nh}_{\mathrm{gal}}}.
\end{align}
Which model is chosen depends on the signal to noise ratio of the
data. Here the partial covering scenario is included in the warm
absorber models.
In addition to the continua above, the Fe K$\alpha$ line at 6.4\,keV
is phenomenologically fitted with a Gaussian model component if
existent. Compton reflection as a more physical model to explain
features like the Fe K$\alpha$ line is not included in the fits
because of the lack of sensitivity above 10\,keV where Compton
reflection dominates. Having found the best fitting model, the
parameter space is further investigated to find the 90\% confidence
levels for each parameter as well as confidence contours between
correlated parameters. In particular correlations between the neutral
column density and the covering fraction as well as the power-law
slope are of interest. Uncertainties of the column density are then
derived from the 90\% confidence contours. The immediate aim is to
search for significant variability of the measured column densities of
a cold absorber that is assumed to be located at the outer torus
layers and hence to underlie structural variability. The significance
of variability is best evaluated by considering contours of all
analyzed observations of one source (Fig.~\ref{fig:combcont}).
The upper panel of Fig.~\ref{fig:combcont} shows that the column
densities measured in different observations are inconsistent between
different epochs, therefore implying a clear detection of variability
in the case of Fairall 51. In contrast, in the case of Mrk~1218
(Fig.~\ref{fig:combcont}, lower panel) the smaller contours of the
{\it XMM-Newton} observation are fully enclosed by the ones of the
{\it Swift} observation with less signal-to-noise. Both observations
are consistent with each other and no variability can be claimed based
on the given data. For models as ours with several degrees of freedom
as well as a limited number of bins, the typical parameter space is
more or less asymmetric \cite{2004physics...6120B}. The results are
asymmetric uncertainties. We derive combined column densities with
unequal upper and lower uncertainties. Gaussian error propagation is
not applicable in this case. A solution is given by
\cite{2004physics...6120B} and also applied in this work.

\section{Results}
\begin{figure}
  \centering
  \includegraphics[width=0.9\columnwidth]{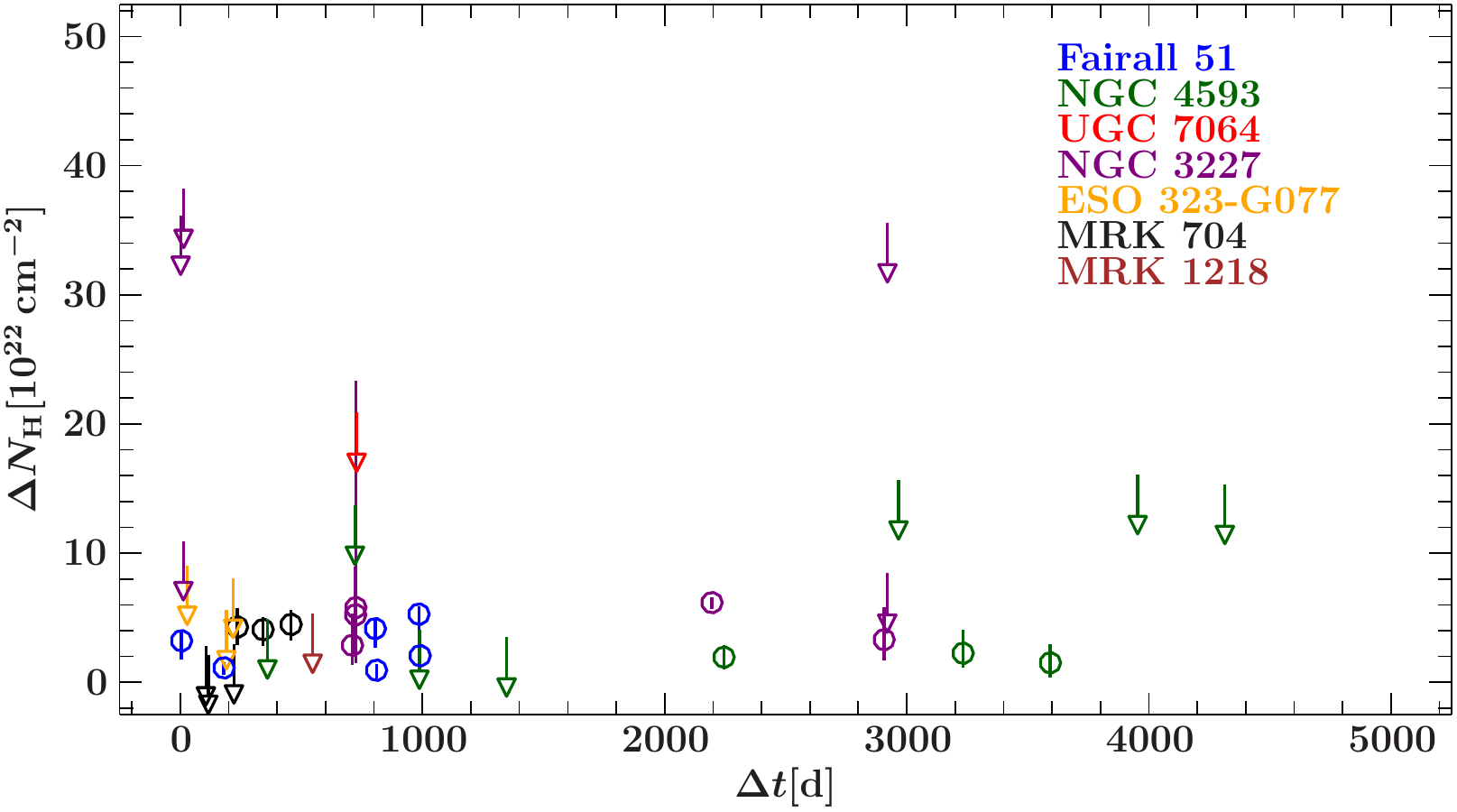}
  \caption{Plot of variability timescales measured between sub-sets of
    two observations against the corresponding difference in
    absorption for all sources with sufficient signal-to-noise ratios. All
    column densities are due to neutral matter except for Fairall~51
    where only warm absorber phases can be constrained.} 
  \label{fig:dt}
\end{figure}

\begin{figure}
  \centering
  \includegraphics[width=0.4\columnwidth]{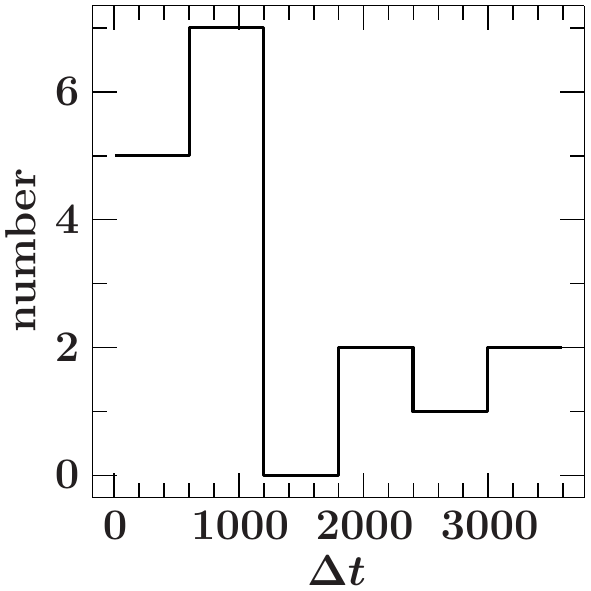}
  \caption{Distribution of all measured variability timescales in days.}
  \label{fig:histo}
\end{figure}

We find 6 out of 12 sources of the sample to reveal variable
absorption of the soft X-rays on timescales from days to years. For
Mrk~766 as a well studied source
\cite{2007A&A...463..131M,2007A&A...475..121T}, minimum variability
timescales on the basis of hours were found by
\cite{2011MNRAS.410.1027R}.
\begin{figure}
  \centering
  \includegraphics[width=0.9\columnwidth]{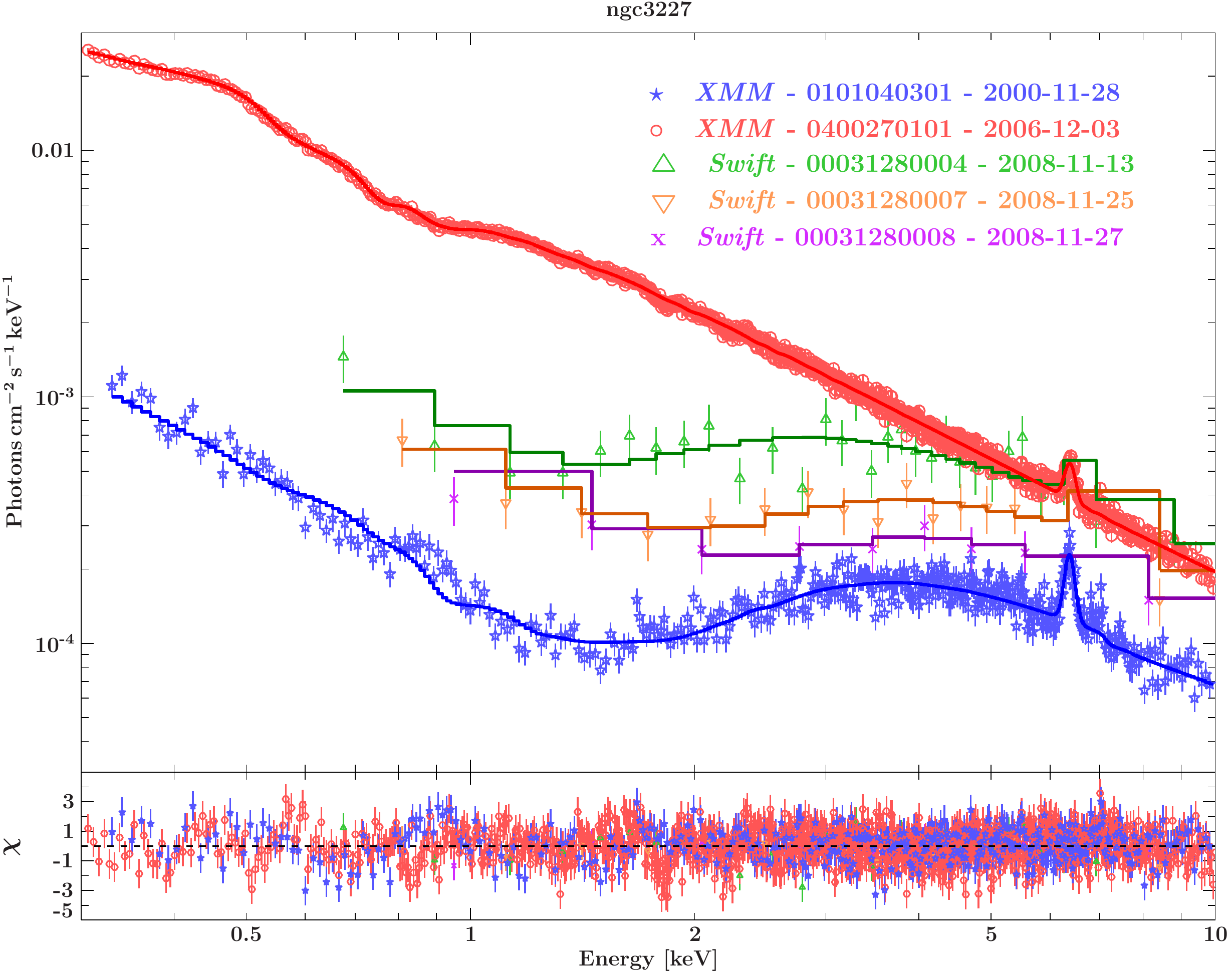}
  \caption{Spectral variability of NGC~3227.}
  \label{fig:ngc3227}
\end{figure}
Figure~\ref{fig:ngc3227} shows an example of strong spectral
variability in NGC~3227 with particularly changing warm absorption
properties. Warm or (partially) gas still has a rich number of
high-$Z$ elements able to absorb soft X-ray photons. In addition to
column density variations of such ionized gas, the ionization states
are also changing between the {\it XMM-Newton} observations. These
changes are interpreted to be due to the varying irradiation expressed
by a varying power-law norm. As the ionization parameter $\xi =
L/n_{\text{e}}r^{2}$ is proportional to the luminosity, it should
behave directly proportional to it. This seems indeed to be the case
for NGC~3227.
Collecting the results from the analysis of the whole sample of
polar-scattered Seyfert\,1 galaxies, Fig.~\ref{fig:dt} shows the time
differences between all possible pairs of observations of all sources
against the appropriate column density differences throughout for all
sources. For the cases where no \nh variability can be stated, upper
limits are shown.
The distribution plotted as histogram is found in Fig.~\ref{fig:histo}. 
Both plots, Fig.~\ref{fig:dt} and Fig.~\ref{fig:histo}, reveal a
strong concentration towards shorter timescales from a few up to
$10^3$\,days. One has to keep in mind that these plots may
represent more the situation of observation timing than a real source
intrinsic distribution of variability timescales. The results of
interest are the shortest timescales of each source as listed in
Table~\ref{tab:mintimescales}. Here we list only sources with
sufficient data to constrain variability.
\begin{table}
\caption{List of minimum timescales for all sources with sufficient
  data to constrain such timescales.} 
\vspace{0.5cm}
\centering
 \begin{tabular}{lll}
\toprule
 Source Name & $\Delta t_{\text{min}}$ \\
\midrule
NGC 3227     & 710\,d \\
NGC 4593     & 2245\,d \\
Mrk 704      & 235\,d \\
Fairall 51   & 5\,d \\
ESO 323-G077 & 455\,d \\
Mrk 766      & $10-20$\,h \cite{2011MNRAS.410.1027R} \\
\bottomrule
 \end{tabular}
\label{tab:mintimescales}
\end{table}

\section{Interpretation}
The shortest variability timescales found are of particular interest
since they trace the smallest spatial scales of the absorber. The
assumed model includes distinct clouds moving across the line-of-sight
on Keplerian orbits. By assuming a certain minimum cloud size to be
able to cover the X-ray source, upper limits on the distance of the
absorber can be derived \cite{2002ApJ...571..234R}. If the radius of
the clouds moving on Keplerian orbits with velocity $v$ around the
central black hole is $r=xr_\mathrm{S}$ where $r_\mathrm{S}=2GM/c^2$
is the Schwarzschild radius, the time for the cloud to pass the line
of sight is
\begin{equation} \label{eq:passtime}
  \Delta t \sim 2r/v = 2xX^{1/2}r_{\text{S}}/c
\end{equation}
with the distance of the absorber $R=Xr_{\text{S}}$
\cite{2012ApJ...749L..31L}. As X-ray sources of AGN are assumed to
have a diameter of $10r_\mathrm{S}$, the cloud diameter must be
$>10r_\mathrm{S}$ in order to cover the central source
\cite{2012ApJ...749L..31L}. The shortest measured timescale of 5\,days
is due to warm absorber variability. Together with a typical black
hole mass of $10^8\,M_\odot$ \cite{padovani1988} we find an absorber
distance of $R \lesssim 1.42\cdot 10^{16}$\,cm. This distance is
consistent with the one to the BLR which is around 0.001--0.1\,pc,
i.e., $3\cdot 10^{15}$--$3\cdot 10^{17}$\,cm away from the black hole
\cite{2005A&A...431..111B,1993PASP..105..247P,1995MNRAS.273.1167R}.
The result also fits well to the model of the BLR to consist of
variable, ionized gas \cite{2005A&A...431..111B,2001ApJ...561..684K}.
The second minimal timescale found for one source is 235\,d in case of
Mrk\,704 (see Table~\ref{tab:mintimescales}). The according upper
limit for the absorber distance estimated to be $R \lesssim 1.1\cdot
10^{19}$\,cm is consistent with the expected distance of the torus of
a few parsec \cite{1988ApJ...329..702K,Pier1992}.

\begin{acknowledgements}
  This work has been funded by the Bundesministerium f\"ur Wirtschaft
  und Technologie under a grant from the Deutsches Zentrum f\"ur Luft-
  und Raumfahrt.
\end{acknowledgements}

\bibliographystyle{actapoly}
\bibliography{mnemonic,aa_abbrv,aaabbrv,literature}

\end{document}